\documentclass[conference]{IEEEtran}
\usepackage{cite}

\ifCLASSINFOpdf
\else
  \usepackage[dvips]{graphicx}
  \graphicspath{{fig/}}
\fi
\usepackage[cmex10]{amsmath}
\usepackage{algorithmic}

\usepackage{mdwlist}
\usepackage[caption=false,font=footnotesize]{subfig}
\usepackage{url}

\usepackage{amssymb}
\usepackage[usenames,dvipsnames]{color}
\usepackage{tabularx}
\usepackage{xspace}
\usepackage{multirow}
\usepackage{multicol}
\usepackage{graphics}
\usepackage{wrapfig}
\usepackage{cite}
\usepackage{algorithm}

 \usepackage{pifont}

\usepackage{pslatex}
\usepackage{mathptmx} 

\usepackage{booktabs} 

\newcounter{tub}
\newcounter{upmc}
\let\oldmarginpar\marginpar
\renewcommand\marginpar[1]{\-\oldmarginpar[\raggedleft\footnotesize
\textcolor{blue}{#1}]%
{\raggedright\footnotesize \textcolor{blue}{#1}}}

\newcounter{mpc}
\newcommand{\xref}[1]{Section~\ref{#1}}
\newcommand{\cref}[1]{Chapter~\ref{#1}}

\newcommand{\fref}[1]{Figure~\ref{#1}}
\newcommand{\tref}[1]{Table~\ref{#1}}
\newcommand{\first}{\emph{(i)}~}
\newcommand{\second}{\emph{(ii)}~}
\newcommand{\third}{\emph{(iii)}~}

\newcommand{\ie}{i.\,e., \@}
\newcommand{\eg}{e.\,g., \@}

\newcommand{\cf}{cf. \@}

\newcommand{\etal}{et~al.\xspace}

\newcommand{\perc}{\,\%\xspace}

\usepackage[normalem]{ulem}
\newcommand{\COMMENTED}[1]{}
\newcommand{\cancel}[1]{}

\definecolor{mygrey}{rgb}{0.8,0.8,0.8}

\newcounter{fn1}
\newcounter{fn2}
\newcounter{fn3}
\newcounter{fn4}
\newcounter{fn5}
\newcommand{\studivz}{\textsc{G-StudiVZ}\xspace}

\newcommand{\twitter}{\textsc{G-Twitter}\xspace}
\newcommand{\synstudivz}[1]{\textsc{G-SynStudiVZ-#1}\xspace}
\newcommand{\syntwitter}[1]{\textsc{G-SynTwitter-#1}\xspace}
\newcommand{\regular}[2]{\textsc{G-Regular-#1-#2}\xspace}

\newcommand{\facebook}{\textsc{P-Facebook}\xspace}
\newcommand{\radius}{\textsc{P-Radius}\xspace}

\renewcommand{\r}[1]{\emph{R#1}}
\newcommand{\m}[1]{\emph{M#1}}

\begin{document}
%
\newcommand{\mytitle}{}
\renewcommand{\mytitle}{Revisiting Content Availability\\in Distributed Online Social Networks}

\title{\mytitle}{}



%
\author{
\IEEEauthorblockN{%
Doris Schi\"oberg\IEEEauthorrefmark{1},
Fabian Schneider\IEEEauthorrefmark{2},
Gilles Tredan\IEEEauthorrefmark{3},
Steve Uhlig\IEEEauthorrefmark{4}, and
Anja Feldmann\IEEEauthorrefmark{1}
}
\IEEEauthorblockA{\IEEEauthorrefmark{1}
\texttt{\{doris|anja\}@net.t-labs.tu-berlin.de} ---
TU Berlin / Telekom Innovation Laboratories, Berlin, Germany
}
\IEEEauthorblockA{\IEEEauthorrefmark{2}
\texttt{fabian@ieee.org} ---
NEC Laboratories Europe, Heidelberg, Germany
}
\IEEEauthorblockA{\IEEEauthorrefmark{3}
\texttt{gtredan@laas.fr} ---
LAAS/CNRS
}
\IEEEauthorblockA{\IEEEauthorrefmark{4}
\texttt{steve@eecs.qmul.ak.uk} ---
Queen Mary University of London, UK
}
}


\maketitle


\begin{abstract}
  Online Social Networks (OSN) are among the most popular applications in
  today's Internet. Decentralized online social networks (DOSNs), a special
  class of OSNs, promise better privacy and autonomy than traditional
  centralized OSNs. However, ensuring availability of content when the content
  owner is not online remains a major challenge.

  In this paper, we rely on the structure of the social graphs underlying DOSN
  for replication. In particular, we propose that friends, who are anyhow
  interested in the content, are used to replicate the users content. We study
  the availability of such natural replication schemes via both theoretical
  analysis as well as simulations based on data from OSN users. We find that
  the availability of the content increases drastically when compared to the
  online time of the user, \eg by a factor of more than 2 for 90\% of the
  users. Thus, with these simple schemes we provide a baseline for any more
  complicated content replication scheme.
\end{abstract}


\section{Introduction}\label{sec:intro}

Online social networks (OSNs) successfully claimed their place among the
most popular Internet services. Despite their success, OSNs controlled by a
single entity raise issues in terms of privacy of content and
communication.  To address these issues, recent works
\cite{buchegger:case,baden:sigcomm09,Diaspora} have proposed decentralized
online social networks (DOSNs), providing privacy and autonomy. 

Privacy of user content involves two different aspects: the access to the
data and the storage/replication of the data. Access to the data can be
restricted through encryption \cite{baden:sigcomm09}, without requiring
trust between the owner of the data and the intermediaries who store it.
Where to store the data in decentralized systems is generally solved by
having an external storage system in charge of keeping replicas of the
data, for example through a distributed file
system~\cite{adya:osdi02,oceanstore:asplos2000}.

The main challenge in decentralization comes from guaranteeing availability
of the data when the owner of the data is not online
\cite{TriblerOverviewJournal}.  Availability has been studied in P2P
file-sharing~\cite{bhagwan:p2ps03,douceur:perfeval03} and distributed file
systems~\cite{adya:osdi02,oceanstore:asplos2000}. 
File sharing is driven by popularity of content instead of social relations.
Most P2P and distributed file systems introduce significant overhead when
replicating data to achieve high availability, without sacrificing high
scalability.

Almost all existing DOSN approaches rely on external storage services and
therefore do not study content availability. 
Those DOSNs which do not rely on external storage amount to exactly one,
i.e. PeerSoN~\cite{schioeberg:peer} which proposes to use storage provided by
a user's friends.
In this paper, we study availability
in such DOSNs that do not rely on external storage
and rather exploit the social graph of a user.  
As direct social friends are interested in a user's content, they
will store it as an effect of their interest. Contrary to existing data
replication schemes that incur computation and communication overhead, our
implicit content replication scheme avoids this overhead.

Given the lack of explicit replication of our strategy, one may expect
that the resulting availability of a user's data is very limited. We show
in this paper that despite the limited replication provided by our
strategy, the availability of content is relatively high. Furthermore, we
show that by allowing a limited fraction of the users to be always online,
\eg by utilizing their own home gateway or an external storage service, the
content availability is comparable to existing OSNs.

This paper makes the following contributions:
\begin{itemize}
\item Using network traces, we study and model the connection
  patterns of today's OSN users.
\item Based on these results, we simulate and analyze the performance of our
  replication scheme. Results show that friend-only replication allows a
  surprisingly good content availability.
\item We study the impact of users being always online and show that
  already a small fraction leads to high overall content availability.
\end{itemize}
    
The remainder of this paper is structured as follows. In \xref{sec:system}
we present our notion of DOSNs, and introduce our replication schemes and
metrics.  In \xref{sec:theory}, we provide an analytical model of the
availability of our schemes. We present our simulation approach and our
models for session characteristics in \xref{sec:models}, before we discuss
our simulation results in \xref{sec:simu}.  We discuss related work in
\xref{sec:related} and conclude in \xref{sec:conc}.

\section{Our DOSN concept and evaluation strategy}\label{sec:system}

We first describe our concept of a Distributed Online Social Network (DOSN)
and define some terminology.  Then we present the content replication
schemes we evaluate in this paper and describe our availability metrics.

\subsection{Terminology and System Description}

DOSN users can engage in social relations with other users, \eg online
friendship, follow another's activity, or subscribe to status updates.
Independent of the type of social relation that connects two users, we call
them \emph{friends} and their relationship \emph{friendship}. The entirety
of users and their relationships form the \emph{social graph} of the DOSN,
where users and relations correspond to nodes and links, respectively.
A user's data---the user's \emph{content}---can be seen as the digital 
representation of a user, which
is stored on a computing device and can be transmitted from one device 
to another. The \emph{content} can contain information on location, work, education,
interests, photos or status updates.
Each user regularly produces such \emph{content}, that she is eager to share with her
friends with the help of the DOSN. We do not make any assumption on the
type of data exchanged by users, but in the rest of the paper, we assume
that the time to transfer a user's data is negligible. Typically content in
OSNs is small in size. Moreover, large objects such as videos can be
uploaded into the cloud (e.g. YouTube) and only the link to this object is
shared through the DOSN. Thus, as we consider DOSN and not P2P file-sharing
systems, this assumption is reasonable.

We consider \emph{DOSN}s in general, without limiting ourselves to a
specific implementation. Yet, we focus on cases where there is no central
server storing users' data, \eg PeerSon~\cite{schioeberg:peer}. 
This implies that availability is a function of users being \emph{online} and
able to serve the data of a given user.

In this paper, we concentrate on DOSN data replication mechanisms.
Therefore, we assume that the typical functions of OSNs, such as finding
online friends and bonding to them, or creating interest groups, is ensured
by another component of the DOSN. 
An often discussed, typical problem in P2P systems is the so called
boot strapping problem, which describes the process and the related issues
when a new node wants to join the system. We will not discuss the boot strapping 
problem in this context because it increases the complexity a lot while giving 
very little insight. 
Note however that our simulation does include nodes with only a few friends.
Further details on boot strapping a DOSN are discussed in~\cite{schioeberg:peer}.

OSNs are highly dynamic, users join and leave, friendships are created and destroyed.
This leads to a very complex scenario. To reduce this complexity we
examine a static snapshot of an OSN. We do not assume 
that the graph is static itself but look at it in a certain state. We
believe this simplification to be reasonable as long as the simulation time
frame is limited. 
Throughout our simulations we do not allow users to join or leave, or
edges to be added or removed. We further assume that for the time
of the simulation the data does not change and because of that stays valid.
In this study we want to follow a piece of data and its distribution over the system
and the resulting availability. If the data would be changed, e.g. by adding new info,
this would be equal to a restart of one of our simulation runs. 

We consider a one-to-one mapping between
a user and a node, and that the node corresponding to a user is always a
replica of this user's data. In other words, nodes always hold a complete
copy of their user's data. We further assume that each user uses exactly one device.
We do not consider the case of one user using
multiple devices (such as a smart-phone and a desktop computer), nor the case
of multiple users sharing the same device (\eg the family's desktop
computer). The assumption that each device holds a full copy is valid, 
because memory is cheap and even smart phones have a lot of storage capacity these
days. 
A full system to handle different versions of data is given 
in~\cite{schioeberg:peer}. This system uses a DHT to store a) information
about available versions of the content and b) locations (\ie which node)
where each version can be found.
This way a user is not necessarily tied to a single device.
Thus, if a user utilizes multiple devices 
to access the DOSN and some device does not hold all content or the most recent
version of the content, the DOSN knows about it. That way friends can make
an informed decision to either download outdated content or wait for fresh
content.
Yet, this case might influence a user's interaction with the DOSN and we
can neither predict nor measure the effects of such a feature. Therefore we
exclude this case from our analysis.
We do not consider shared devices for two reasons. Either they only share data 
when the user is online, then there is not differences to our ``exactly-one-device''-rule. 
Another option is that the device shares the data of all its users 
while any of the users is online. For that case our assumptions actually 
constitute a worst case.

\subsection{Replication schemes}
\label{sec:replication-schemes}

\begin{figure}
\includegraphics[width=\linewidth]{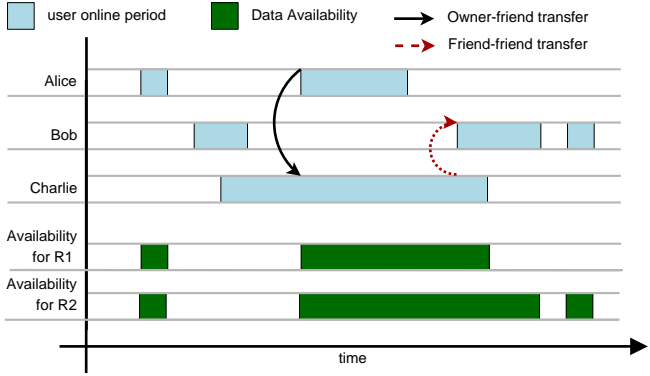}
\caption{Our replication strategy for user Alice. Bob and Charlie are friends of
  Alice. Once two friends (\eg Alice and Charlie, first arrow) are
  simultaneously online they exchange their data: Charlie gets Alice's data. In
  the replication scheme \r1, Bob never gets Alice's data since they are never
  online together. In replication scheme \r2, Bob downloads Alice's data from
  Charlie (second arrow), which leads to a better availability.}
\label{fig:replication}
\end{figure}

To improve data availability, data can be replicated on multiple nodes. These
nodes become \emph{replicas}. Choosing good replicas is a crucial question
that has received a lot of attention in the context of distributed file sharing. 
In the context of DOSN however, one structure is by definition available: the 
underlying social graph. In this paper, we build an implicit replication strategy
from this underlying social graph. The rationale behind this strategy is the fact
that social friends are natural candidates for replicating a user's data, as they
are interested in that person. And since memory is cheap, friends can store
a replica of an item even if they are not interested in it.
We also assume that each user takes care to backup her own data. 
In case a friend's hardware fails, the user still has the full data set and can
transmit it to her friend as soon as their hardware works again.
We study three different replication mechanisms that constitute different ways 
to exploit the social graph of a DOSN for data replication. The example in 
\fref{fig:replication} illustrates these replication schemes for user
Alice:

\begin{description}  

\item[\r0] \emph{No replication:} In this scheme, only the user provides
her own data. In other words, the data is available only when the
corresponding user is online. It is arguably not a replication mechanism,
but constitutes a baseline for comparison of the other schemes. In
\fref{fig:replication}, Alice's \r0 availability is equal to Alice's online
time. This is the only case where it could be a difference if a user 
has multiple devices. If Alice is online with her mobile phone and adds a new data
item, this item will not be available in that moment from her home
PC. 
On the other side, in that moment the home PC would probably not be online. In
case it is, synchronization mechanisms like those explained in~\cite{schioeberg:peer} will
take care that both devices reach the same state again.

\item[\r1] \emph{Direct replication only:} In this scheme, the data is made
available by the user and her friends. To be able to distribute user's
data, friends must obtain a copy of this data directly from the user
itself. In \fref{fig:replication}, Alice's \r1 availability is her online
time, plus the online time of Charlie after he got a copy of Alice's data.
Since Bob and Alice were never simultaneously online, Bob cannot replicate
Alice's data.

\item[\r2] \emph{Indirect replication:} In this scheme, friends of a
user can collaborate with other friends to obtain this user's data when
they are online.  In \fref{fig:replication}, Bob can in this case get
Alice's data from Charlie, and distribute it. Alice's \r2 availability is
thus made by Alice, Charlie after he got the copy from Alice, and Bob after
he got the copy from Charlie.

\end{description}

Note that if we represent the social network by its adjacency matrix $A$,
the links of $A^0=Id, A^1=A$, and $A^2$ correspond to replication graphs of
\r0, \r1, and \r2 strategies respectively, where an edge in the replication
graph represents data exchange. It is however important to recall that data
is only hosted by direct friends: common friends of a user might exchange
the data of this common friend, but do not host each other's data. For
instance in \fref{fig:replication} if Bob is not Charlies' friend, he will
get Alice's data from Charlie, but will not host Charlie's data.  Note also
that to keep this ``only direct friends are replicas rule'', strategy \r2
is the best we can do.

\subsection{Metrics: Pure availability and Friend Availability}
\label{sec:metr-pure-avail}

The term \emph{availability} can have many definitions in the context of
P2P and distributed systems. Definitions include hardware failure rate,
churn rate of peers, or information about which parts of content can be
downloaded from which peer.  In this study we consider content
availability, \ie which fraction of time the data of a user is available to
others in a DOSN. The content is available when the user's node is online,
or when a replica of the data is online.

One option to measure the availability would be to count how many data
requests are successfully answered in a real system. This would require to
model and simulate user requests. Although this is possible, we refrain
from doing so because it substantially increases the simulation complexity
and duration. Moreover, it is difficult to find real-world data that
includes data request behavior and derive a good model from it. Therefore,
we rely on time-based availability metrics.

We do not consider data updates in this study. Instead we assume that each
user generates new content at some point of the simulation. We measure
availability over a day (24 hours) following the new content generation.
Note, that this also means that the only copy of the data
available in the system at the beginning is held by the user's node. We
believe this constitutes a worst-case scenario. Consider
\fref{fig:replication}: The first time Bob is online he still has not
downloaded Alice's data. We therefore consider Alice's data as unavailable.
In reality, Bob has probably a local copy of Alice's older data that he can
serve as well: he just does not have the latest update.

Here, we do consider two different types of availability. The first one is
the traditional content availability, as defined above: The fraction
of time a piece of data is available. We refer to it as \emph{pure
availability} or metric \m1.

In the context of DOSN however, we can again exploit the social graph. People
interested in a given user's data are primarily her friends. Recall that
any kind of OSN relationship defines friendship. In a news service,
only the followers that subscribed to a feed will receive the content. In 
Facebook or Google Plus many profiles are only shared with friends, 
such that an arbitrary user cannot see the profile.

Thus, in this paper, we also measure the \emph{friend availability} or
metric \m2, which is the fraction of time a user's data is available 
when her friends are online.  This
availability takes into account who is interested in the data that is
available. This last availability can be seen as a \emph{convergence metric}
as once all the friends of a user have a copy of the data, this friend
availability is one, even though friends are rarely online. Note also that pure
availability can be higher than friend availability. For instance 
assume \fref{fig:replication} represents one day and Bob is Alice's only
friend. Because Alice and Bob are never simultaneously online, \m2 would be
zero but \m1 corresponds to Alice's availability.

In the remainder of this paper we refer to the combination of the replication
scheme \r{x} and the metric \m{y} as \r{x}\m{y}, \eg the combination of no
replication and pure availability is referred to as \r0\m1 and indirect
replication and friend availability is referred to as \r2\m2.

\section{Theoretical analysis of availability}
\label{sec:theory}

Data availability has already been theoretically addressed in the context of
peer-to-peer storage
\cite{Weatherspoon02erasurecoding}.  Despite
the similarity of making data available in peer-to-peer networks and
distributed replication of OSN user data, these problems differ on some
fundamental points such as the model of replica failures.

\subsection{Model and assumptions}\label{sec:toward-an-analytical}

Let $G(V,E)$ be the graph modeling the OSN friendship connectivity. We assume a
discrete time model, and that $G$ does not vary over time. Let $\alpha_i(t)$ be
the probability that node $i$ is online at time $t$. We assume that all nodes
have the same session characteristics:  $\forall t, \forall i \in V,
\alpha_i(t)=\alpha(t)$. The graph dynamics is captured by nodes being offline or
online. We also assume that the probability of two nodes being connected
(online) at a given time $t$ are independent of time: $P(\textit{``i and j are
  online at time t''})=\alpha_i(t)\alpha_j(t)=\alpha(t)^2$. In other words
connections are independent and identically distributed. In the following,
$\overline{a}$ denotes the complement of $a$.

Most of the related work focuses on achieving a highly reliable data storage
system despite failures or departures of the peers holding the data. For
instance in \cite{ramabhadran2006analysis}, a dynamic model is studied where
peers continuously fail, and the replication rate must be adapted to compensate
for these losses. Our model on the other hand considers that friends are
reliable: they never fail nor depart from the system. Once the user data has
been replicated to all its friends, the system is converged.
The main aspect of our model is to consider that nodes can be
temporarily disconnected by being dynamically online and offline.

\subsubsection{Data availability metric}\label{sec:metr-data-avail}

The goal of P2P storage systems is to store large amounts of data in a reliable
fashion, at low cost. Considerations such as the available disk space of a replica 
or the available bandwidth are therefore important. For instance, some of the works 
on P2P systems consider the use of error correcting codes as a mean to reduce the 
consumed space while keeping the desired availability. In this work, we 
assume that enough storage is available at every node.

In this paper, we are interested in the user's data availability: during which fraction of 
time is user $i$'s data available? Formally, let $A_i$ be this availability, and
$\Omega$ be the measuring period: $$A_i = \frac{1}{\vert \Omega \vert} \sum_{k
  \in \Omega} P( i\textit{'s data is available at time } k).$$

Given that replication is made only among directly connected neighbors, the problem 
we address is a \textit{local} one with respect to $G$. As we will see, the most relevant 
graph property in our context is the out-degree of nodes, \ie the number of potential 
replicas. Other graph properties such as diameter, connectedness or clustering do not 
have a major impact on availability.

\subsubsection{Converged case}\label{sec:conv-case}

We consider that a given node $i$ is in a \textit{converged} state when all its
neighbors have obtained a copy of $i$'s data, and are therefore able to act as
$i$'s data replica. This constitutes an upper bound of the availability $A_i$
since $i$ cannot expect more replicas. Let $n$ be the number of neighbors of
node $i$. With the assumption of independence between online times, we
have $$A_i\leq P(\overline{i\textit{'s replicas are all offline}}) = 1- (1 -
\alpha(t))^{n+1}.$$

This is the maximal availability a node with $n$ neighbors can hope for in our
model.

\begin{figure}
\centering
\includegraphics[width=\linewidth]{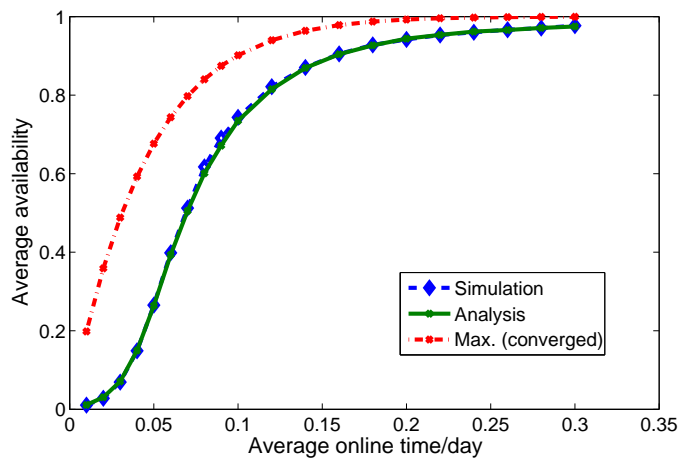}
\caption{Match between simulations and analysis} 
\label{fig:theoryplot}
\end{figure}

\begin{figure}
\centering
\vspace{0.8\baselineskip}
\includegraphics[width=\linewidth]{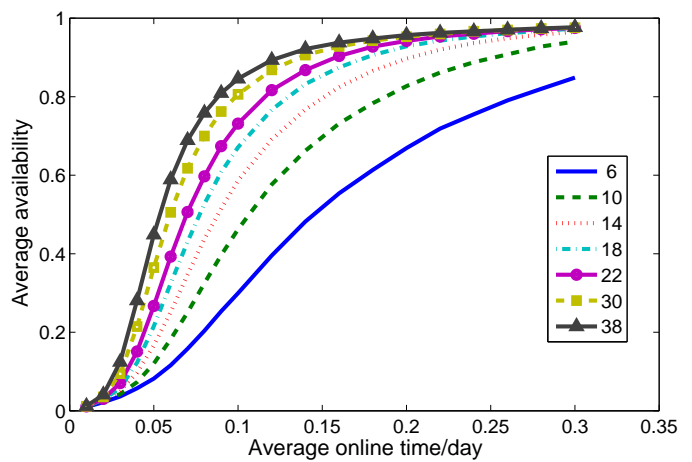}
\caption{Impact of the degree on average pure availability (analysis).}
\label{fig:theoryplot2}
\end{figure}

\subsubsection{Non-converged case}
\label{sec:noconv-case}

Fix node $i$. Let $N_i(t)$ be the random variable representing the number of $i$'s
replica in the system at time $t$ (be they online or not). We have
$P(N_i(0)=1)=1$, since the only available copy of $i$'s data at time $0$ is on
node $i$. Note that $N_i(t)$ is strictly increasing: no replica ever deletes a
hosted user's data. Now, assume there are $k-j$ replicas in the system at time
$t-1$. Then the probability that $N_i(t)=k$ is the probability that exactly
$j$ nodes became replicas at time $t$. In other words, using total probability law:
\begin{eqnarray}
  P(N_i(t)=k)= & \sum_{j=0}^{k-1} P(N_i(t-1)=k-j) \cdot \nonumber\\&
   P(X_t=j\vert N_i(t-1)=k-j) \label{eq:Ni}.
\end{eqnarray}
Where $X_t$ is the random variable representing the number of replications
happening at time $t$. Let us express $P(X_t=j| N_i(t-1)=k-j)$, the probability
that exactly $j$ new replicas appear at time $t$.  Note that we need to
differentiate the case where no new replica is created at this time ($j=0$) for
the general case. Assume $j=0$, two reasons for this: either no replica
connected, or at least one replica was online, but no non-replica friend
connected. Thus it writes:
\begin{eqnarray*}\footnotesize
&  P(X_t=0| N_i(t-1)=k) \\
=&(1-\alpha(t))^k + (1-\alpha(t))^{n-k}\sum_{p=1}^{k}
  {k-j \choose p} \alpha(t)^p(1-\alpha(t))^{k-p}\\
  =& (1-\alpha(t))^k + (1-\alpha(t))^{n-k}(1-(1-\alpha(t))^{k})
\end{eqnarray*}

Now assume $j>0$, at least one replica is created, which also means that at
least one replica is online, and exactly $j$ friends that are still not
replicas connect. In the following,
using total probability law, we decompose according to the number $p$ of
replicas that are online at time $t$, and use the fact that if at least one
replica is online, the number of new replicas created is exactly the number of
non-replica friends that are online:
\begin{eqnarray*}
\footnotesize
& P(X_t=j| N_i(t-1)=k-j) =\\
 & \sum_{p=1}^{k-j}P(X_t=j \text{ and p online replicas}\vert N_i(t-1)=k-j) \\
  =& \sum_{p=1}^{k-j}P(\text{ p online replicas}\vert N_i(t-1)=k-j).\\
  &P(X_t=j\vert\text{ p online replicas and } N_i(t-1)=k-j)\\
  =& \sum_{p=1}^{k-j} {k-j \choose p} \alpha(t)^p(1-\alpha(t))^{k-j-p} . {n-k+j
    \choose j}\alpha(t)^j(1-\alpha(t))^{n-k} \\
 =&  (1-(1-\alpha(t))^{k-j}){n-k+j
    \choose j}\alpha(t)^j(1-\alpha(t))^{n-k} 
\end{eqnarray*}
Since we know that $P(N(0)=1)=1$ we can then iteratively compute $P(N(t)=k)$.

Now that we can describe the evolution of $i$'s number of replicas over time, we
can express the probability that $i$'s data is available at time $t$: it is
the complement of finding all the available replicas offline. Let $A(t)$ be
the event ``$i$'s data is available at time $t$''. We have
\begin{equation}
  \label{eq:5}
  P(A(t))=\sum_{k=0}^n P(N_i(t)=k)(1-(1-\alpha(t))^k).
\end{equation}

Using this relation, it is possible to numerically estimate the average
availability of a given node over time with Monte-Carlo simulation (code
is available\cite{gilles-code}). Figure~\ref{fig:theoryplot}
compares the analytically derived average availability over a day with the
results of the simulation, for different online patterns. We consider here nodes
with a degree $d=22$, and the probability of a node to be online in a given time
bin is drawn uniformly at random over the day ($\alpha(t)=\alpha$). The
theoretical maximum availability, \ie the availability achieved in the converged
case, is also plotted. The gap between the maximal availability and both
analysis and simulations of convergence decreases as the average online time
increases. This confirms the intuition that the more often nodes are online, the
faster the converged situation is reached, \ie all potential replicas have a
copy of the data.

Figure \ref{fig:theoryplot2} presents the average availability of nodes as a
function of their average online time for various node degrees. The point
($x=0.2$,$y=0.65$) on the lower (blue) curve shows that according to our
analysis, nodes with $6$ neighbors being online with $20\%$ probability in each
time bin have on average $65\%$ availability. This illustrates the strong impact
of node degree on availability: nodes with $30$ neighbors need to be online
nearly 3 times less to achieve the same availability, and can expect more than
$97\%$ availability with $20\%$ online time.

\section{Simulation methodology}\label{sec:models}

We evaluate DOSN content availability for different graphs and user online
patterns. In this section, we describe the friendship graphs used, explain
how we model the session characteristics of DOSN users, and present the
simulation procedure.

\subsection{Social Graphs and their Node Degree Distribution}
\label{sec:graphs}

\begin{table}
\centering
\caption{Summary over relationship graphs}
\label{tab:graphs}
\begin{tabular}{lrrl}
\toprule
Name 			& \#Nodes 	& avg. 		&  \\
			& 		& Degree	& Distribution \\
\midrule
\studivz		&  1.04~M	& 22.24		& \multirow{2}{*}{Weibull (0.9; 22.5)}\\
\synstudivz{$N$}	&    $N$	& 24.12		& \\
\cmidrule(rl){1-4}
\twitter		& 51.22~M 	& 41,71		& 
\multirow{2}{*}{Power-law (2.25; 41)}\\
\syntwitter{$N$}	&    $N$ 	& 41		& \\
\cmidrule(rl){1-4}
\regular{$N$}{$D$}	&    $N$ 	& $D$	& $N$ nodes w/ degree $D$\\
\bottomrule
\end{tabular}
\end{table}

As shown in \xref{sec:theory},
the availability of content depends strongly on the social graph,
especially its node degree distribution. For this study we use relationship
graphs from two well-known OSNs, as well as synthetically generated graphs.
The synthetic graphs allow us to explore social graphs with different
average node degrees as well as the influence of graph size. \tref{tab:graphs}
summarizes the relationship graphs we use in this study.

Regarding the real-world graphs, we use graphs derived from crawls by
Fritsch~\cite{studivz-graph} for the \studivz graph
and Cha \etal~\cite{icwsm10cha} for the \twitter graph.
StudiVZ is a popular Facebook like OSN for German speaking students.
While \studivz is
symmetric due to reciprocal friendships, \twitter is asymmetric. Being a
follower is a unidirectional relation. 

The Twitter dataset includes a link from user $A$ to user $B$, when the
crawl revealed that $A$ follows $B$. This notion of edge direction is
contrary to ours. We consider edges to point in the direction in which
content is transferred: From the content originator to the friend that is
interested in the data. In Twitter content flows from the followed ($B$) to
the follower ($A$). Therefore, when we want to use the Twitter dataset we
need to consider the distribution of in-degrees, instead of out-degrees.

Both of these graphs have more than a million nodes, \twitter has
50~million. Running simulations with that many nodes is not feasible when
exploring all our combinations of simulation parameters. In order to reduce the
computation time we synthetically generate smaller graphs (\eg $100,000$
nodes) that follow the same node degree distribution as the real-world
graphs, using the \texttt{gengraph} tool by Viger and
Latapy~\cite{viger05graphgen}.

\begin{table*}
\centering
\caption{Summary over session characteristics}
\label{tab:onoff}
\begin{tabular}{lcccccccc}
\toprule
Name 		& \multicolumn{5}{c}{Durations}
		& \multicolumn{2}{c}{Arrivals/bin } 		&
		\multirow{2}{5em}{\centering Sessions\\per day}\\
\cmidrule(rl){2-6}
\cmidrule(rl){7-8}
		& Q1 	& median& mean 	& Q3 	& fitted Model
		& low 		& high 		& \\
\midrule
\facebook	& 0m37s	& 6m30s	& 69m	& 52m	& Weibull (0.4; 1284) $[s]$
		& -97\perc	& +107\perc	& 2.5\\
		  
\radius w/o always-on 	
		& 3m10s & 5m16s	& 50m & 19m	& Weibull (0.35; 550) $[s]$
		& -52\perc	& +44\perc 	& 4.5 \\
\bottomrule
\end{tabular}
\end{table*}

\begin{figure}
\includegraphics[angle=-90,width=\linewidth]{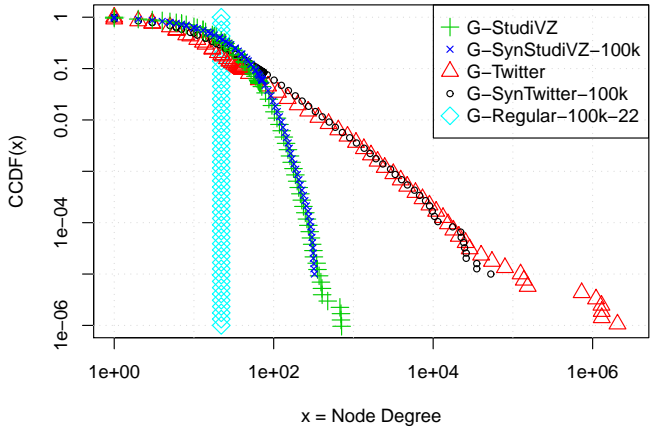}
\caption{CCDF of node degree distributions of the graphs used. Axis are
cutoff at 100.000 nodes for increase comparability with the syntetic
graphs.}
\label{fig:graphs-nd}
\end{figure}

To produce the input degree distributions for \texttt{gengraph}, we fit the
\studivz degrees with a Weibull distribution with shape 0.9 and scale 22.5.
For Twitter's in-degrees we fit a power-law distribution with $\alpha=2.25$
and a mean of 41, using the node degree distribution generator
\texttt{distrib}, which is part of \texttt{gengraph}.  We validate our
fitting through Kolmogorov-Smirnov (KS) tests as well as visual inspection
(\fref{fig:graphs-nd}).  For comparison, we also generate regular graphs, \ie
graphs where each node has the same degree, with \texttt{gengraph}. We use
these graphs to study the impact of different (average) node degrees and in
contrast to the heavy-tailed real-world distributions.

In our simulations we do not consider that the underlying social graphs are
changing. However, as users go online and offline the graph of active users
is highly dynamic. We characterize and model these changes in the following
subsection.

\subsection{Session characteristics}
\label{sec:onoff}

For our time-based simulations, it is important to reproduce the session
characteristics of DOSN users. Although we were able to get access to
relationship graphs from real-world OSNs, there are no publicly available
models of session durations and arrivals for those graphs. Therefore, we
rely on two types of session characteristics derived from data gathered in
the aggregation network of a large European ISP.  First, we consider
session characteristics of Facebook sessions observed from about 2,000
users (\facebook), see Schneider \etal~\cite{schneider09imc} for details on
the dataset and a study of popular user interactions with OSNs.  Second, we
consider DSL session characteristics (\radius) from about 20,000 customers,
see Maier \etal~\cite{maier09imc} for details on the dataset.

From our DSL session data we observed that a significant fraction (57\perc)
of lines are always connected, \ie they perform an automatic reconnect upon
disconnects from the ISP. The exact fraction heavily depends on
the (default) configuration of the DSL router, and is subject to change
depending on the set of services provided by the ISP, \eg a VoIP customer
DSL line should be always connected. We choose to filter out those sessions
and instead use a parameter in the simulations denoting the fraction of
always-on nodes.

Because the number of users for which we observed the session
characteristics does not match with the sizes of our social graphs, we need
to determine a session model that can scale with the number of users.
For each session we identify the login and logout timestamps and model
the session \emph{start times} and \emph{durations}. Our session start time
model accounts for time of day effects. We model the arrivals using a
modulated Poisson process with 20-minute bins. In each bin the rate  
is modulated according to the observed probability that a session is
starting in that bin.

We did not observe a strong correlation between the session's start time
(of day) and its duration. Pearson's correlation coefficient is $-0.28$ and
$-0.06$ for \facebook and \radius, respectively. Therefore, we model the
session durations independently from the session start times, using Weibull
distributions. Again we validate the fittings using KS tests. 

\tref{tab:onoff} summarizes the session characteristics observed. We show
the mean durations as well as the median, together with the fitted
distributions.  Note that for \radius the statistics and the model only
correspond to those DSL lines which are not permanently connected (referred
to as "w/o always-on").  In addition, we show the deviation from the
average number of arrivals per bin and number of sessions per user per day.

\subsection{Simulation setup}

Our simulation takes three parameters: \first the social graph, \second the
session characteristics, and \third a configuration that determines the
duration of the experiment and the time period considered to compute the
availability metrics.

First we select a social graph which in turn defines $N$, the number of
nodes/users in the DOSN, and the node degree distribution as described in
\xref{sec:graphs}.

Second, we select one of the two session models from \xref{sec:onoff}, and
generate a (possibly empty) set of presence times for each of the $N$ nodes
of the graph. A presence time consists of the start and end time of an
online session. In addition to the session model, the online times of users
depend on the fraction $P$ of nodes that are always connected.
Note that the session models include time-of-day effects, that is nodes are
online and offline
depending on the time of day. Further the process used to generate the 
sessions randomly applies different sessions to all users, so that
two users do not have the same on/off-pattern.

We generate a total of $(1-P) \times N \times \varnothing$
presence times, where $\varnothing$ is the average number of sessions per
day and user (see \tref{tab:onoff}). This allows to compare the results for
graphs with different numbers of nodes. The session start times follow
the time-of-day modulated bins. The session end times are produced by adding
a duration drawn from the Weibull distribution corresponding to the session
model. To keep simulation runs with the same session model comparable, we
generate one big set of session durations per model and use them subsequently. 
Each session lasts for at least 5 seconds.

For the remaining $P \times N$ nodes, we set the session start time to the
beginning of the simulation and the session end time to the end of the
simulation time. Once everything is prepared we compute the availability
of each user's content for the different replication schemes and availability
metrics in one pass.

\section{Simulation results}\label{sec:simu}

In this section, we start by demonstrating that our reduced size
graphs accurately capture availability. Then, we study the availability
that can be expected from our socially-driven replication schemes, and
highlight the importance of the way the availability metric is defined. We
also discuss the impact of different social graphs and user session models.
We close this section by assessing the improvements in availability gained
from a fraction of always-online nodes as well as considering longer
simulation periods.

\begin{figure}
\includegraphics[angle=270,width=\linewidth]{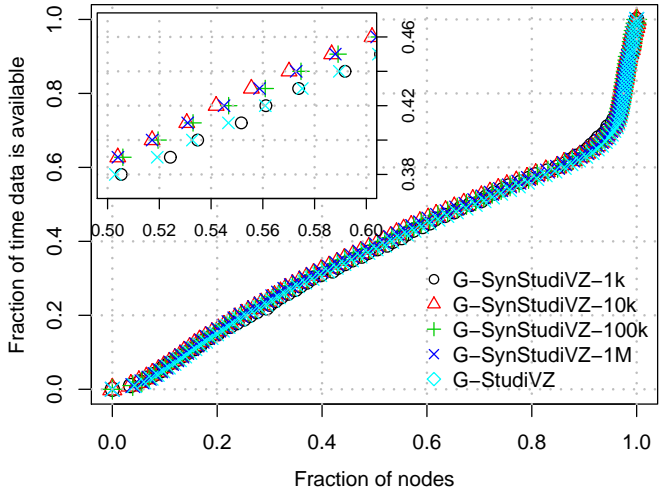}
\caption{Comparing \studivz and synthetic graphs of varying size, using
\facebook and \r2\m1: Using small synthetic graphs is viable.}
\label{fig:res-size}
\end{figure}

\subsection{Graph sizes do not matter}

In the remainder of this paper, we present results as cumulative distribution
function (CDF) of the content availability (y-axis) as a function of the
fraction of nodes/users (x-axis).

\fref{fig:res-size} shows the availability of \studivz based graphs with a
\facebook session model over 24\,h simulation time for \r2\m1. Before
computing the CDFs, we remove all nodes from the simulation output which
never go online, and therefore never generate any content. For all simulations 
this exclusion affects around 6\perc of the nodes for \facebook and 1\perc for 
\radius.

Since simulating the replication process is an expensive task in terms of
computational time, we do most of the experiments on scaled-down graphs.
In \xref{sec:graphs} we already showed the match between synthetic graphs
and real-world graphs in terms of node degree distribution. In
\fref{fig:res-size} we now show that also the resulting availability is
very similar, \cf \studivz and \synstudivz{1M}. Moreover, scaling down
the size of the network does not affect the availability distribution
either, \cf \synstudivz{1k}, \synstudivz{10k} and \synstudivz{100k}.

Based on this insight, we use the synthetic graphs with 100,000 nodes
for all further simulations. Using smaller graphs significantly speeds
up simulation time and thereby allows to explore a broader variety and
combination of parameters. Unless otherwise mentioned we present results 
for \synstudivz{100k} and \facebook for 24\,h, which serves as our baseline
scenario.

\begin{figure}
\includegraphics[angle=270,width=\linewidth]{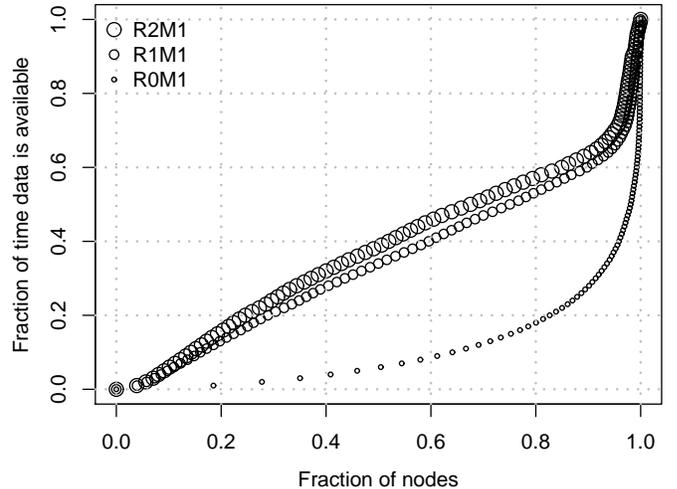}
\caption{Major gains in availability using replication schemes (\r2,\r1) over no
replication (\r0) for \facebook. \r1 is only marginally worse than \r2.}
\label{fig:res-replication}
\end{figure}

\subsection{The Gain of Replication}

Next, we turn to understanding the impact of the different replication
schemes that we introduced in \xref{sec:replication-schemes}.
\fref{fig:res-replication} shows \r0\m1, \r1\m1 and \r2\m1 for our baseline
scenario. As compared to \r0\m1 which represents the availability of the
nodes themselves (no replication), we observe a drastic increase in
availability for both replication schemes \r1 and \r2. While \r2 offers
limited gains in availability for the bottom and top 5\perc of nodes ranked
by \r0, the vast majority (90\perc) of nodes double their availability and
the median availability is shifted from less than $5\%$ to more that
$35\%$---an increase by a factor of 7.  The bottom 5\perc of nodes suffer
from the reduced opportunities to hand over their data, due to their own
limited online time. The top 5\perc do not gain much as their own online
time is already high.

Taking a detailed look at \fref{fig:res-replication}, we see that
replication schemes \emph{R2} and \emph{R1} produce very similar results.
As expected, \r2 has better availability compared to \r1. However, the
similar results indicate that passing along content over multiple friends
does not happen often. One explanation for this is that contacts with the
data owner occur before contacts between two friends. Indeed, increasing
the simulation time to multiple days did not change this result.

Note that due to the small difference between \r1 and \r2 and given that
the restriction of allowing to copy the data only from the originator is
artificial, \eg when relying on encryption of content, we refrain from
presenting the results for \r1 in the sequel. Further, \r0 always produces
the same curve for identical presence times, which only varies for
different graph sizes and session characteristics.

\begin{figure}
\includegraphics[angle=270,width=\linewidth]{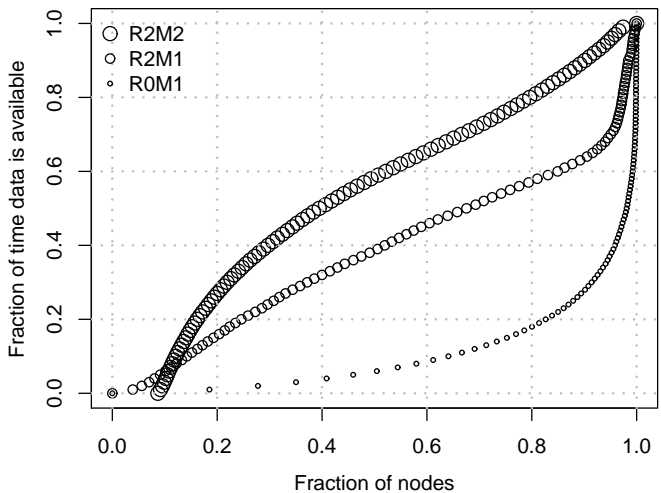}
\caption{Availabilty of content with respect to total friends online time (\m2)
vs. total simulation time (\m1): \m2 yields higher availability.}
\label{fig:res-metrics}
\end{figure}

\subsection{Difference between availability metrics}

So far, we looked at the availability measured in relation to the total time
of the experiment, which we defined in \xref{sec:system} as our
first metric \m1.  Our second metric \m2 is the data availability measured
in relation to the online time of a user's friends.  In
Figure~\ref{fig:res-metrics} we observe that \r2\m2 is a significant
improvement over \r2\m1 for the baseline scenario. The median availability
increases to around 60\perc. This corresponds to a 30\perc gain for more
than 60\perc of the nodes over \r2\m1.

However, the availability of about 8\perc of nodes reduces to zero for
\r2\m2, \cf bottom left part of plot. Indeed, if no friend is online at the 
same time as the node itself, \m2 reports 0 while \m1 reports the node's 
own availability. Our finding from the previous subsection holds, in that 
\r2 is slightly better than \r1 when comparing \r1\m2 with \r2\m2 (not 
shown). Without replication, \m2 is still better than \m1 but by very little. 
Therefore, from now on we focus on \r0\m1 (as reference), as well as 
\r2\m1 and \r2\m2 whenever useful.

\begin{figure}
\includegraphics[angle=270,width=\linewidth]{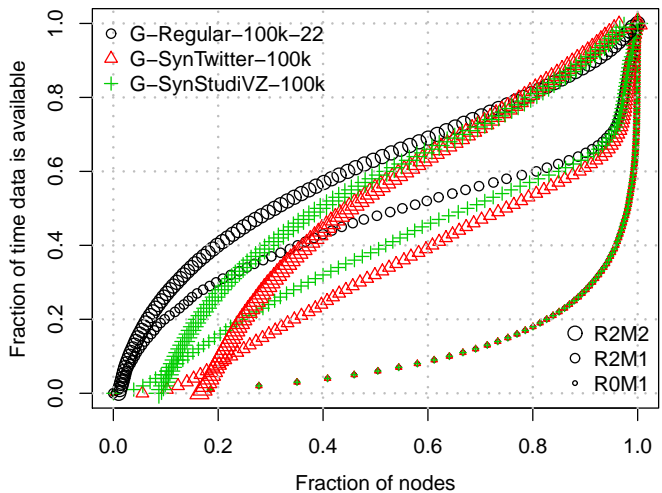}
\caption{Impact of graph type on availability, using \facebook}
\label{fig:g-types}
\end{figure}

\subsection{Impact of graphs types and node degree}

As introduced in \xref{sec:graphs}, we use different types of graphs
with different node degree distributions. In Figure~\ref{fig:g-types}, we
present the availability of two graphs in addition to our \synstudivz{100k}
baseline graph: \syntwitter{100k} and \regular{100k}{22}. We choose the
average node degree of \studivz (22) as node degree for the regular graphs,
to allow comparison.

\regular{100k}{22} achieves the best availability, followed by
\synstudivz{100k}.  \syntwitter{100k} has the worst availability, although
it has the highest average node degree. From this observation, we conclude 
that the degree distribution is more important than the average degree for
availability.  Both synthetic graphs have a lot of nodes with only a few friends, 
so that their opportunities to get their data replicated is lower. Indeed,
\syntwitter{100k} has significantly more nodes with single digit node
degrees than \synstudivz{100k} (50\perc over 35\perc).
The \regular{100k}{22} graph does not have nodes with zero \r2\m2
availability, as nodes do not need to rely on the opportunity to meet their 
one and only friend.

\begin{figure}
\includegraphics[angle=270,width=\linewidth]{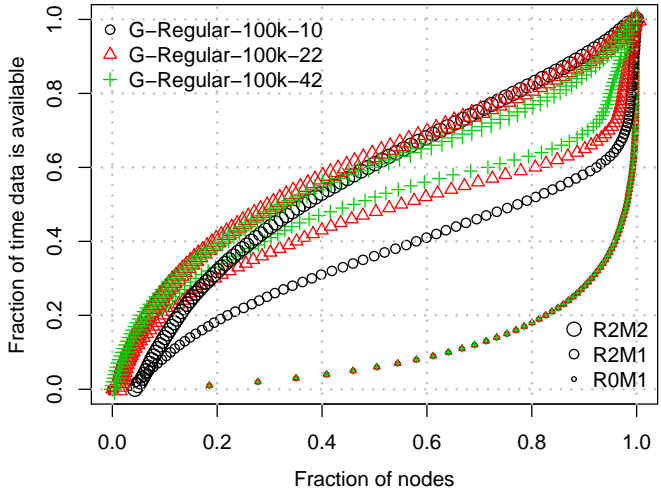}
\caption{Impact of node degree on availability, using \facebook}
\label{fig:g-degrees}
\end{figure}

To study the impact of a graph's node degree, we consider different
versions of \regular{100k}{?}.  Besides the degree of $22$ we also used
\regular{100k}{10}, and \regular{100k}{42}. Figure~\ref{fig:g-degrees} 
shows that the degree of each individual node has an impact on the 
availability, although it is limited, as can be observed through the saturation 
effect around 22 neighbors. While \r2\m1 is always better for higher node 
degrees, \r2\m2 is slightly worse for the top half of nodes (right sides of plot).

\begin{figure}
\includegraphics[angle=270,width=\linewidth]{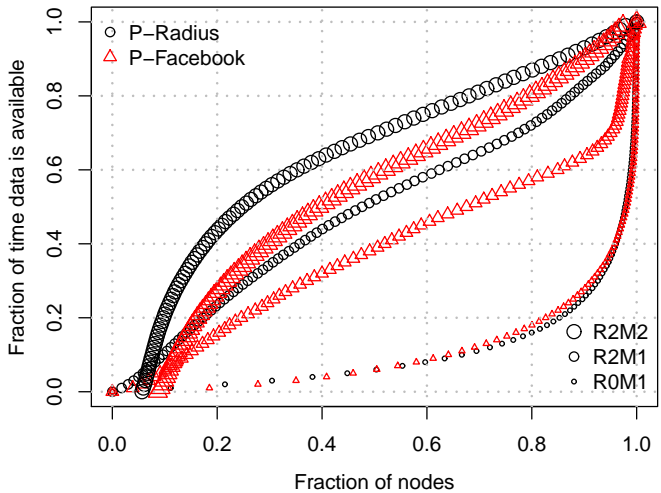}
\caption{Impact of session characteristics on availability, using
\synstudivz{100k}}
\label{fig:res-arrivals}
\end{figure}

\subsection{Impact of session characteristics}

As described in \xref{sec:onoff}, we use mainly two different
distributions for the session characteristics, \radius and \facebook. 
Figure~\ref{fig:res-arrivals} compares those on \studivz. \radius shows
higher availability for \r2\m1 and \r2\m2. Yet, for \r0\m1 the difference
is limited. However, the gap between the two session models is certainly 
bigger than the gap between different graphs, highlighting the higher
importance of the session models compared to the graph.

The difference in availability for the replication cases, corresponds to
the expected total online time of all nodes.  As shown in \tref{tab:onoff},
the product of session per day and average duration is higher for \radius.
This matches the expectation that a user usually is more often and longer
online in the Internet than logged in to Facebook. Obviously, a user first
needs to establish an Internet access connection before using Facebook.

The reason why the higher total online time of \radius is not as pronounced
in the \r0\m1 curve is that the median session durations are tiny compared 
to the simulation duration, \eg 6m30s corresponds to a value of 0.0045 on 
the y-axis.

\begin{figure}
\includegraphics[angle=270,width=\linewidth]{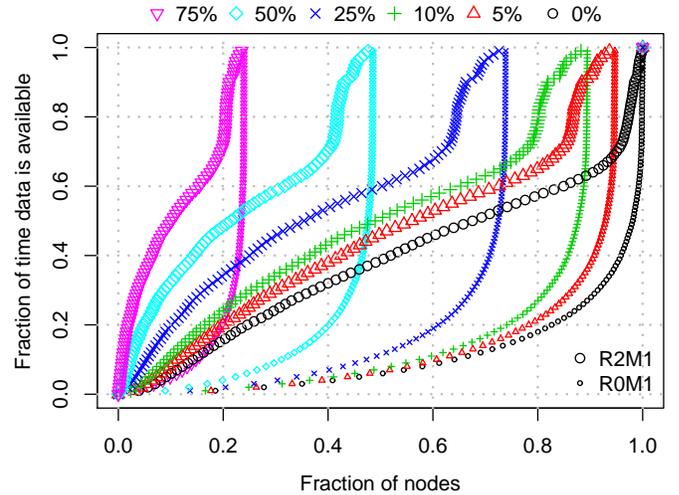}
\caption{Availability with $X$\perc always on nodes.} 
\label{fig:res-alwayson}
\end{figure}

\subsection{Adding always on nodes}

Many DOSNs rely on external storage and consider storage as always
available. Similarly, we saw more than half (57\perc) of all \radius
sessions being connected all the time. Therefore we now explore how much 
we can improve availability by replacing a fraction $P$ of nodes with always-on
nodes. This can be achieved by running the DOSN client on a high
availability platform, e.g., their home gateway which is usually always
online, instead of their computer.

\fref{fig:res-alwayson} (legend on top) shows our baseline scenario for
$P=\{0,5,10,25,50,75\}\%$. Note, the compressed curves due to this 
fraction of always available nodes. Replacing only 5\perc of the 
nodes, already increases the \r2\m1 availability of the 10\perc top most
available nodes to almost full availability.

\subsection{Considering longer time-frames}

\begin{figure}
\includegraphics[angle=270,width=\linewidth]{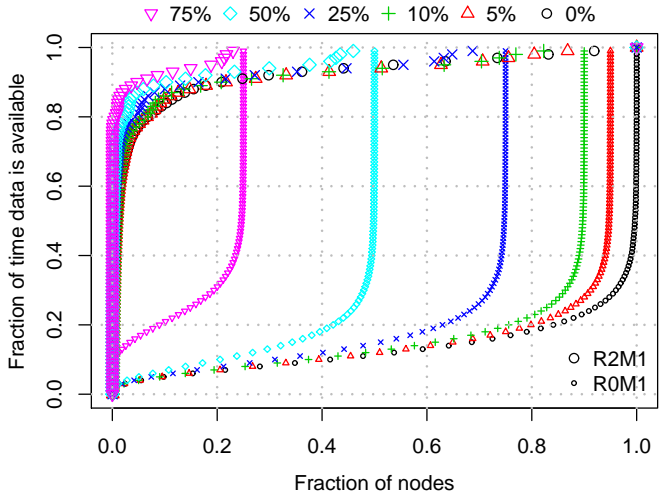}
\caption{Always on nodes for 7\,days simulation.}
\label{fig:res-avail7days}
\end{figure}

If in addition to always-on nodes, we consider a longer time frame until
the data is requested, more contacts will have occurred and the content is
spread to more replicas.  Considering a time frame of a full week instead
of 24\,h, friends (\r2\m2) have almost a 100\perc guarantee to obtain the
desired data, as shown in \fref{fig:res-avail7days}. In the most extreme
case, looking at a 7~day experiment with 75\perc always on nodes, each nodes
availability is at least 80\perc.

\begin{figure}
\includegraphics[angle=270,width=\linewidth]{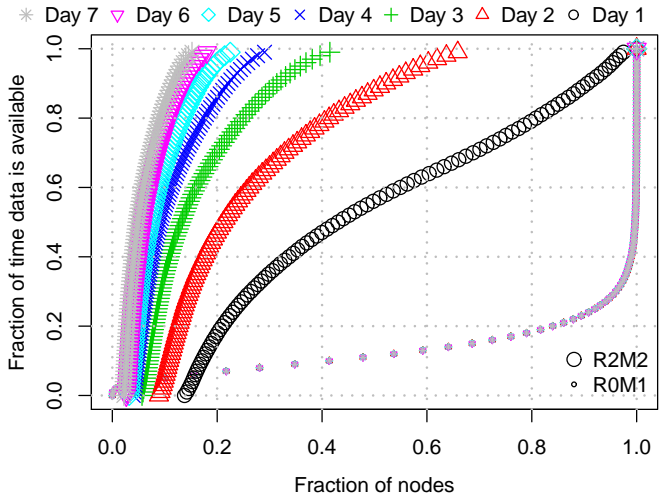}
\caption{Daily availability after each of the 7\,days.}
\label{fig:res-availTime}
\end{figure}

Observing such high availability we are interested in determining the
convergence towards this full weeks results. Therefore, in 
\fref{fig:res-availTime} we plot the \r2\m2 availability of DOSN
contents for each day of the week. We observe steady but decreasing
improvement on each day. Culminating in an almost vertical curve, at the
last day of the week more than 80\perc of all the nodes have full
availability.

\section{Related work}
\label{sec:related}

Two major strategies have been studied for data replication in distributed systems: 
replicate the whole file or replicate the pieces of a file. 
Works such as \cite{PAST-Hotos,Bhagwan02replicationstrategies} are examples of the first
strategy. PAST~\cite{PAST-Hotos} is a P2P file system built on top of Pastry~\cite{Pastry-Middleware}, 
a DHT-based system. \cite{Bhagwan02replicationstrategies} explores different variations of the gain of 
replicating data to randomly chosen hosts on data availability. \cite{Bhagwan02replicationstrategies}
shows that implicit replication that follows file popularity, as in Gnutella and BitTorrent, is not sufficient.
The second replication approach splits a file into a pre-defined number of pieces and
distributes them across the nodes.
Examples of this approach include \eg \cite{DBLP:journals/corr/abs-0803-0632}, or
\cite{DBLP:journals/corr/abs-1008-0064}. \cite{DBLP:journals/corr/abs-1004-4438}
extensively covers the related work in this area. Replicating the whole file and pieces of a file
can be combined, as done in OceanStore~\cite{oceanstore:asplos2000} that uses a combination 
of different replication strategies for different data types, \eg  erasure codes for archiving.
\cite{Weatherspoon02erasurecoding} compares the two replication strategies and finds that
erasure codes lead to better availability than redundancy.

Availability has been studied in the context of P2P networks. \cite{p2pavial:QinXin} explores 
the properties of different coding schemes and finds that schemes which give better availability 
are in general the ones that have higher maintenance cost. \cite{bhagwan:p2ps03} measures 
and analyzes host availability in a large structured P2P file sharing network. They find that host 
availability is dependent on the time of day, but not on other hosts. \cite{Chu02availabilityand} 
confirmed the results of \cite{bhagwan:p2ps03} in Gnutella and Napster. \cite{Saroiu02ameasurement} 
also explore Gnutella and Napster but focus on the nature of end-hosts, and find they are very 
heterogeneous, \eg regarding access bandwidth and their limited cooperation.

DOSNs have gained popularity in the research community in the recent years, aimed at giving back
users control over their data. Most DOSN projects concentrate on the question of access rights 
management and encryption. They rely on servers or other external services to guarantee data 
availability. Others use social links for trust but do not exploit it for DOSN data availability. 
Tribler~\cite{TriblerOverviewJournal} is a file sharing system built on top of a social overlay. It 
uses a complex replication system to ensure the best possible availability of data. Persona~\cite{baden:sigcomm09} 
relies on external storage services that can be anything from a server to an Amazon cloud account 
per user. Persona enables privacy by introducing a fine grained system that users can use to manage 
access to their data.
Diaspora~\cite{Diaspora} recently obtained much attention from the press.
From the users viewpoint, 
Diaspora appears to be a common server-based OSN. However, data can be encrypted and everyone 
can set up a server for Diaspora, so that availability is ensured by many distributed servers. 
SuperNova~\cite{DBLP:journals/corr/abs-1105-0074} is an architecture for a DOSN that solves the 
availability issue by relying on super-peers that provide highly available storage.

\vspace{\baselineskip}

\section{Conclusion}\label{sec:conc}

In this paper, we proposed to exploit on the structure of the social graphs in 
DOSNs, to replicate users' content. As the friends of a given user are interested 
in his content anyway, they can be used to provide replicas of his content. Through 
theoretical analysis as well as simulations based on data from OSN users, we study
this natural replication scheme. We find that with such a replication scheme, the 
availability of the users content increases drastically, when compared to the online 
time of the users, \eg by a factor of more than 2 for 90\perc of the users. Furthermore,
adding a small fraction of always online users to our scheme leads to high overall
availability of users' content. 

As future work, we want to study the speed at which user content updates 
propagate through the graph of DOSNs, as well as compare our replication
scheme with others that are not purely based on the friendship graph.

We also need to leave studies to future work that look at the overhead in
terms of traffic, time, messaging generated by a system similar to the
presented one. Such a system would need to be implemented, deployed and
observed while in use to find out about those details.

\vspace{\baselineskip}





\bibliographystyle{IEEEtran}
\bibliography{paper}

\begin{thebibliography}{10}
\providecommand{\url}[1]{#1}
\csname url@samestyle\endcsname
\providecommand{\newblock}{\relax}
\providecommand{\bibinfo}[2]{#2}
\providecommand{\BIBentrySTDinterwordspacing}{\spaceskip=0pt\relax}
\providecommand{\BIBentryALTinterwordstretchfactor}{4}
\providecommand{\BIBentryALTinterwordspacing}{\spaceskip=\fontdimen2\font plus
\BIBentryALTinterwordstretchfactor\fontdimen3\font minus
  \fontdimen4\font\relax}
\providecommand{\BIBforeignlanguage}[2]{{%
\expandafter\ifx\csname l@#1\endcsname\relax
\typeout{** WARNING: IEEEtran.bst: No hyphenation pattern has been}%
\typeout{** loaded for the language `#1'. Using the pattern for}%
\typeout{** the default language instead.}%
\else
\language=\csname l@#1\endcsname
\fi
#2}}
\providecommand{\BIBdecl}{\relax}
\BIBdecl

\bibitem{buchegger:case}
S.~Buchegger and A.~Datta, ``A case for {P2P} infrastructure for social
  networks - opportunities \& challenges,'' in \emph{Proceedings of the Sixth
  international conference on Wireless On-Demand Network Systems and Services
  (WONS'09)}, 2009, pp. 149--156.

\bibitem{baden:sigcomm09}
R.~Baden, A.~Bender, N.~Spring, B.~Bhattacharjee, and D.~Starin, ``Persona: an
  online social network with user-defined privacy,'' in \emph{Proceedings of
  the ACM SIGCOMM 2009 conference on Data communication (SIGCOMM '09)}, 2009,
  pp. 135--146.

\bibitem{Diaspora}
D.~Grippi, M.~Salzberg, R.~Sofaer, and I.~Zhitomirskiy, ``Diaspora,''
  \url{https://joindiaspora.com/}.

\bibitem{adya:osdi02}
A.~Adya, W.~J. Bolosky, M.~Castro, G.~Cermak, R.~Chaiken, J.~R. Douceur,
  J.~Howell, J.~R. Lorch, M.~Theimer, and R.~P. Wattenhofer, ``Farsite:
  federated, available, and reliable storage for an incompletely trusted
  environment,'' in \emph{Proceedings of the 5th symposium on Operating systems
  design and implementation (OSDI '02)}, 2002, pp. 1--14.

\bibitem{oceanstore:asplos2000}
J.~Kubiatowicz, D.~Bindel, Y.~Chen, S.~Czerwinski, P.~Eaton, D.~Geels,
  R.~Gummadi, S.~Rhea, H.~Weatherspoon, W.~Weimer, C.~Wells, and B.~Zhao,
  ``{OceanStore}: An architecture for global-scale persistent storage,'' in
  \emph{Proceeedings of the Ninth International Conference on Architectural
  Support for Programming Languages and Operating Systems (ASPLOS'00)}, 2000,
  pp. 190--201.

\bibitem{TriblerOverviewJournal}
J.~Pouwelse, P.~Garbacki, J.~Wang, A.~Bakker, J.~Yang, A.~Iosup, D.~Epema,
  M.~Reinders, M.~van Steen, and H.~Sips, ``Tribler: A social-based
  peer-to-peer system,'' \emph{Concurrency and Computation: Practice and
  Experience}, vol.~20, pp. 127--138, 2008.

\bibitem{bhagwan:p2ps03}
R.~Bhagwan, S.~Savage, and G.~M. Voelker, ``Understanding availability,'' in
  \emph{Proceedings of the Second International Workshop on Peer-to-Peer
  Systems (IPTPS'03)}, 2003, pp. 256--267.

\bibitem{douceur:perfeval03}
J.~R. Douceur, ``Is remote host availability governed by a universal law?''
  \emph{SIGMETRICS Performance Evaluation Review}, vol.~31, pp. 25--29, 2003.

\bibitem{schioeberg:peer}
D.~Schi{\"{o}}berg, ``A peer-to-peer infrastructure for social networks,''
  Diplomarbeit, TU Berlin, Germany, 2008.

\bibitem{Weatherspoon02erasurecoding}
H.~Weatherspoon and J.~Kubiatowicz, ``Erasure coding vs. replication: A
  quantitative comparison,'' in \emph{In Proceedings of the First International
  Workshop on Peer-to-Peer Systems (IPTPS'01)}, 2002, pp. 328--338.

\bibitem{ramabhadran2006analysis}
S.~Ramabhadran and J.~Pasquale, ``Analysis of long-running replicated
  systems,'' in \emph{Proceedings of the 25th IEEE International Conference on
  Computer Communications (INFOCOM'06)}, 2006, pp. 1 --9.

\bibitem{gilles-code}
G.~Tredan, \url{http://www.net.t-labs.tu-berlin.de/~gilles/availability.tgz}.

\bibitem{studivz-graph}
H.~Fritsch, ``Studivz - {I}noffizielle {S}tatistiken vom {D}ezember 2006,''
  \url{http://studivz.irgendwo.org/}.

\bibitem{icwsm10cha}
M.~Cha, H.~Haddadi, F.~Benevenuto, and K.~P. Gummadi, ``Measuring user
  influence in twitter: The million follower fallacy,'' in \emph{Proceedings of
  the 4th International AAAI Conference on Weblogs and Social Media
  (ICWSM'10)}, 2010.

\bibitem{viger05graphgen}
F.~Viger and M.~Latapy, ``Efficient and simple generation of random simple
  connected graphs with prescribed degree sequence,'' in \emph{Computing and
  Combinatorics, LNCS 3595}, 2005, \texttt{gengraph} is available at
  \url{http://www-rp.lip6.fr/~latapy/FV/generation.html}.

\bibitem{schneider09imc}
F.~Schneider, A.~Feldmann, B.~Krishnamurthy, and W.~Willinger, ``Understanding
  online social network usage from a network perspective,'' in
  \emph{Proceedings of the 9th ACM SIGCOMM conference on Internet measurement
  (IMC'09)}, 2009, pp. 35--48.

\bibitem{maier09imc}
G.~Maier, A.~Feldmann, V.~Paxson, and M.~Allman, ``On dominant characteristics
  of residential broadband internet traffic,'' in \emph{Proceedings of the 9th
  ACM SIGCOMM conference on Internet measurement (IMC'09)}, 2009, pp. 90--102.

\bibitem{PAST-Hotos}
P.~Druschel and A.~Rowstron, ``Past: A large-scale, persistent peer-to-peer
  storage utility,'' in \emph{Proceedings of the Eighth Workshop on Hot Topics
  in Operating Systems (HotOS VIII)}, 2001, pp. 75--.

\bibitem{Bhagwan02replicationstrategies}
R.~Bhagwan, D.~Moore, S.~Savage, and G.~M. Voelker, ``Replication strategies
  for highly available peer-to-peer storage,'' in \emph{Future directions in
  distributed computing}, 2003, pp. 153--158.

\bibitem{Pastry-Middleware}
A.~Rowstron and P.~Druschel, ``Pastry: {Scalable,} decentralized object
  location and routing for large-scale peer-to-peer systems,'' in
  \emph{Proceedings of the IFIP/ACM International Conference on Distributed
  Systems Platforms (Middleware)}, 2001, pp. 329--350.

\bibitem{DBLP:journals/corr/abs-0803-0632}
A.~G. Dimakis, B.~Godfrey, Y.~Wu, M.~J. Wainwright, and K.~Ramchandran,
  ``Network coding for distributed storage systems,'' \emph{CoRR}, vol.
  abs/0803.0632, 2008.

\bibitem{DBLP:journals/corr/abs-1008-0064}
F.~E. Oggier and A.~Datta, ``Self-repairing homomorphic codes for distributed
  storage systems,'' \emph{CoRR}, vol. abs/1008.0064, 2010.

\bibitem{DBLP:journals/corr/abs-1004-4438}
A.~G. Dimakis, K.~Ramchandran, Y.~Wu, and C.~Suh, ``A survey on network codes
  for distributed storage,'' \emph{CoRR}, vol. abs/1004.4438, 2010.

\bibitem{p2pavial:QinXin}
Q.~Xin, T.~Schwarz, and E.~L. Miller, ``Availability in global peer-to-peer
  storage systems,'' in \emph{Distributed Data and Structures 6, Proceedings in
  Informatics}, 2004.

\bibitem{Chu02availabilityand}
J.~Chu, K.~Labonte, and B.~N. Levine, ``Availability and locality measurements
  of peer-to-peer file systems,'' in \emph{Proceedings of ITCom: Scalability
  and Traffic Control in IP Networks}, 2002.

\bibitem{Saroiu02ameasurement}
S.~Saroiu, P.~K. Gummadi, and S.~D. Gribble, ``A measurement study of
  peer-to-peer file sharing systems,'' in \emph{Proceedings of the Multimedia
  Computing and Networking (MMCN)}, 2002.

\bibitem{DBLP:journals/corr/abs-1105-0074}
R.~Sharma and A.~Datta, ``Supernova: Super-peers based architecture for
  decentralized online social networks,'' \emph{CoRR}, vol. abs/1105.0074,
  2011.

\end{thebibliography}

%
%

\end{document}